\documentclass[aps,prb,twocolumn,showpacs]{revtex4-1}

\usepackage{graphicx}
\usepackage{calc}
\usepackage{bm}
\usepackage{color}
\usepackage{textcomp}

\begin{document}

\title{Lateral Josephson effect on the surface of Co$_3$Sn$_2$S$_2$ magnetic Weyl semimetal}

\author{O.O.~Shvetsov}
\author{V.D.~Esin}
\author{Yu.S.~Barash}
\author{A.V.~Timonina}
\author{N.N.~Kolesnikov}
\author{E.V.~Deviatov}
\affiliation{Institute of Solid State Physics of the Russian Academy of Sciences, Chernogolovka, Moscow District, 2 Academician Ossipyan str., 142432 Russia}

\date{\today}

\begin{abstract}
 We experimentally study lateral electron transport between two 5~$\mu$m spaced superconducting indium 
leads on a top of magnetic Weyl semimetal Co$_3$Sn$_2$S$_2$. For the disordered magnetic state of 
Co$_3$Sn$_2$S$_2$ crystal, we only observe the Andreev reflection in the proximity of each of the leads, which
is indicative of highly transparent In-Co$_3$Sn$_2$S$_2$ interfaces. If the sample is homogeneously 
magnetized, it demonstrates well-developed anomalous Hall effect state. In this regime we find the
Josephson current that takes place even for 5~$\mu$m long junctions and shows the unusual magnetic field and 
temperature dependencies. As a possible reason for the results obtained, we discuss the contribution
to the proximity-induced spin-triplet Josephson current from the topologically
protected Fermi-arc states on the surface of Co$_3$Sn$_2$S$_2$.
\end{abstract}


\maketitle

\section{Introduction}
    
   Similarly to topological insulators~\cite{hasan} and quantum Hall systems~\cite{buttiker,ESreview}, Weyl
semimetals~\cite{mag1,armitage} (WSM) have topologically protected  surface states. They are Fermi arcs 
connecting projections of Weyl nodes on the surface Brillouin zone and these surface states 
inherit the chiral property of the Chern insulator edge states~\cite{armitage}. WSMs should have either
space-inversion or time-inversion symmetry to be broken. First experimentally investigated WSMs were 
non-centrosymmetric crystals, where spin- and angle- resolved
photoemission spectroscopy  data have demonstrated spin-polarized surface Fermi arcs~\cite{das16,feng2016}.

There are only a few candidates of magnetically ordered materials~\cite{mag1,mag2,mag3,mag4} for the 
realization of WSMs in the  systems with broken time reversal symmetry. Recently, giant anomalous Hall 
effect (AHE) was reported~\cite{kagome,kagome1} for the kagome-lattice ferromagnet Co$_3$Sn$_2$S$_2$ as an 
anomalous Hall conductance in zero magnetic field. The AHE can be regarded as the indication to a  
magnetic Weyl phase~\cite{armitage}, as supported by the topological-insulator-multilayer model, where the 
two-dimensional Chern edge states form the three-dimensional WSM surface states 
\cite{BurkovBalents}. The Fermi arc surface states, indeed, were directly 
visualized in Co$_3$Sn$_2$S$_2$ by scanning tunneling spectroscopy~\cite{kagome_arcs}.

For normal metals, Andreev reflection~\cite{andreev} allows charge transport from the metal (N) to 
superconductor (S)  through the NS interface by creating  a Cooper pair at energies below the superconducting
gap~\cite{andreev,tinkham}. For two closely spaced superconducting leads, i.e. for the SNS junction, multiple 
Andreev reflection can contribute to the subharmonic structure of the current-voltage 
characteristics~\cite{tinkham,agrait}.

For topological materials, the proximity to a superconductor usually demonstrates non-trivial 
physics~\cite{Fu,ingasb,nbwte}. As it was experimentally shown, the Josephson coupling in a topological 
insulator is established through the surface conducting channels~\cite{hyeong}. The edge current contribution
can be retrieved even for systems with conducting bulk by analyzing the Josephson current 
behavior~\cite{yakoby,kowen,inwte}.  In  Weyl semimetals, various topological superconducting states can 
appear~\cite{corr1,corr2,corr3,blockade} and various types of Andreev reflection can take place 
\cite{spec,blockade}, depending on the particular conditions. Thus the specular Andreev reflection, 
reminiscent the one in the graphene ~\cite{been1,been2}, can take place at the Weyl semimetal - Weyl
superconductor interface~\cite{spec}, while the chirality blockade of Andreev reflection can appear at the 
interface of the magnetic Weyl semimetal and the conventional s-wave spin-singlet superconductor~\cite{blockade}.

For magnetically ordered topological materials the proximity to a superconductor is a new and emerging 
field involving the mutual influence of superconductivity and magnetism  under nontrivial topological
conditions, e.g., in the presence of topologically protected interface states. For example, it has been 
theoretically identified that the proximity to a superconductor can result in the Majorana modes originating
from the Fermi arc in a Weyl semimetal wire with an axial magnetization \cite{Majorana1}. It has also been
predicted~\cite{toki}, that the proximity induced superconducting surface states of magnetically doped 
topological insulators can represent chiral Majorana modes. Thus, it is reasonable to study  proximity effects
in superconducting junctions, fabricated on a three-dimensional magnetic Weyl semimetal surface.

Here, we experimentally study lateral electron transport between two 5~$\mu$m spaced superconducting indium 
leads on a top of magnetic Weyl semimetal Co$_3$Sn$_2$S$_2$. For the disordered magnetic state of 
Co$_3$Sn$_2$S$_2$ crystal, we only observe the Andreev reflection in the proximity of each of the leads, which
is indicative of highly transparent In-Co$_3$Sn$_2$S$_2$ interfaces. If the sample is homogeneously 
magnetized, it demonstrates well-developed anomalous Hall effect state. In this regime we find the
Josephson current that takes place even for 5~$\mu$m long junctions and shows the unusual magnetic field and 
temperature dependencies. As a possible reason for the results obtained, we discuss the contribution
to the proximity-induced spin-triplet Josephson current from the topologically
protected Fermi-arc states on the surface of Co$_3$Sn$_2$S$_2$.

\section{Samples and technique}

\begin{figure}
    \includegraphics[width=\columnwidth]{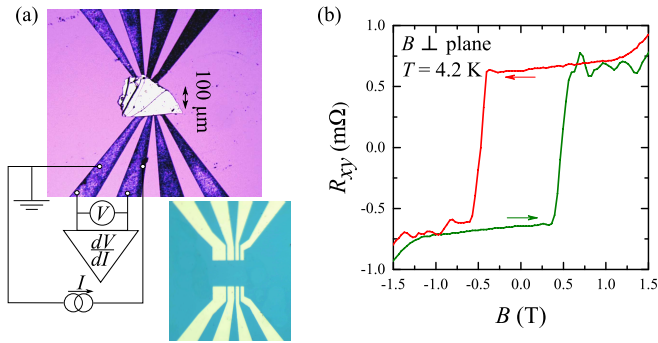}
    \caption{(Color online) (a) A top-view image of the sample with the sketch of electrical connections. A flat (about 100~$\mu$m size and 1~$\mu$m thick) single-crystal Co$_3$Sn$_2$S$_2$ flake is weakly pressed on the insulating SiO$_2$ substrate with 100~nm thick, 5~$\mu$m separated  In leads. The leads pattern is demonstrated in the bottom image. Non-linear $dV/dI(I)$ curves are measured by a standard four-point technique, all  the wire resistances are excluded. (b)   Giant anomalous Hall effect, which confirms high quality of our  Co$_3$Sn$_2$S$_2$ samples~\cite{kagome,kagome1}. Arrows indicate the field scanning directions. }
    \label{cosns_sample}
\end{figure}

Co$_3$Sn$_2$S$_2$ single crystals were grown by the gradient freezing method. Initial load of high-purity 
elements taken in stoichiometric ratio was slowly heated up to 920$^\circ$~C in the horizontally positioned 
evacuated silica ampule, held for 20 h and then cooled with the furnace to the ambient temperature at the rate
of 20 deg/h. The obtained ingot was cleaved in the middle part. The Laue patterns confirm the hexagonal 
structure with $(0001)$ as cleavage plane.  Electron probe microanalysis of cleaved surfaces and X-ray 
diffractometry of powdered samples confirmed stoichiometric composition of the crystal.

Weyl semimetals are essentially  three-dimensional~\cite{armitage} macroscopic crystals. Despite it is possible to form contacts directly on the  polished  crystal plane, we use another technique, which is known to provide highly transparent contacts~\cite{inwte,nbwte,cdas}. The leads 
pattern is formed on the insulating SiO$_2$ substrate by lift-off technique after thermal evaporation of 
100~nm In, see Fig.~\ref{cosns_sample}~(a). The indium leads are separated by 5~$\mu$m intervals. Since the 
kagome-lattice ferromagnet Co$_3$Sn$_2$S$_2$ can be easily cleaved along (0001) crystal plane, small (about 
100~$\mu$m size and 1~$\mu$m thick) Co$_3$Sn$_2$S$_2$ flakes are obtained by a mechanical cleaving method. 
Then we select the most plane-parallel flakes  with clean surface, where no  surface defects could be resolved
with optical microscope. They are transferred to the In leads pattern and pressed slightly with another 
oxidized silicon substrate. A special metallic frame allows us to keep the substrates parallel and apply a 
weak pressure to the sample. No external pressure is needed for a Co$_3$Sn$_2$S$_2$ flake to hold on to a 
substrate with In leads afterward. This procedure provides transparent contacts, stable in different cooling 
cycles, which has been also demonstrated before~\cite{inwte,nbwte,cdas}.

Magnetoresistance measurements confirms high quality of the prepared  Co$_3$Sn$_2$S$_2$ samples. We check that
samples  demonstrate giant anomalous Hall effect, as it has been previously reported~\cite{kagome,kagome1} for
Co$_3$Sn$_2$S$_2$ semimetal. Fig.~\ref{cosns_sample}~(b) shows hysteresis behavior and sharp switchings in 
Hall resistance $R_{xy}$, the switchings' positions $\approx 0.5$~T even quantitatively coincide with the 
reported values~\cite{kagome,kagome1}. According to our sample dimensions,  the Hall resistivity $\rho_{xy}$ can be estimated as 1~$\mu\Omega$cm in zero magnetic field, which is three times smaller in comparison with  Refs.~\cite{kagome,kagome1}.

We study electron transport between two 5~$\mu$m separated In leads by a standard four-point technique. The 
principal circuit diagram is depicted in Fig.~\ref{cosns_sample}~(a). In this connection scheme, all  the 
wire resistances are excluded, which is necessary for low-impedance  In-Co$_3$Sn$_2$S$_2$-In junctions.  To obtain $dV/dI(I)$ characteristics, the dc current $I$ 
(up to 1~mA)  is additionally modulated by a low ($\approx$5~$\mu$A) ac component. We measure both dc ($V$) 
and ac (which is proportional to $dV/dI$) components of the voltage drop  with a dc voltmeter and a lock-in, respectively, after 
a broad-band preamplifier. The measurements are performed in a dilution refrigerator for the temperature 
interval 30~mK--1.2~K.

If the SNS junction demonstrates zero resistance, an important information can be obtained~\cite{yakoby,kowen,inwte} from the maximum supercurrent $I_c$ suppression by temperature $T$ and magnetic field $B$. To obtain $I_c$ values with high accuracy for given $(B,T)$ values,  we sweep current $I$ ten times from zero value (superconducting  $dV/dI=0$ state) to some value well above the $I_c$ (i.e. to the resistive $dV/dI>0$ state), and then determine  $I_c$ as an average value of $dV/dI=0$ breakdown positions in different sweeps.

\section{Experimental results}

\begin{figure}
    \includegraphics[width=\columnwidth]{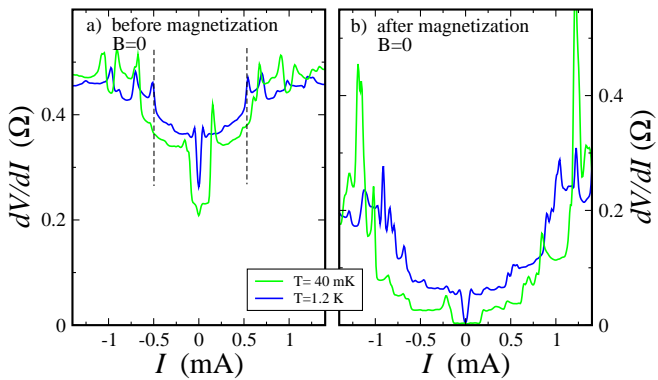}
    \caption{(Color online) Examples of $dV/dI(I)$ curves before sample magnetization (a) and after it  (b), for the same sample in a single cooling cycle. Diminishing the temperature from 1.2~K to 40~mK has low effect on the initial Andreev-like  $dV/dI(I)$ curves in  (a). Dashed lines  indicate the wide $dV/dI$ resistance drop in (a), caused by Andreev reflection.  In contrast, the magnetization procedure  transforms $dV/dI(I)$ curves into the Josephson-like ones in (b). The width of a wide zero-resistance region depends on the temperature below 1.2~K. The data are obtained in zero magnetic field.}
    \label{IV}
\end{figure}

Co$_3$Sn$_2$S$_2$ magnetic properties arise from the kagome-lattice cobalt planes, whose magnetic moments 
order ferromagnetically~\cite{kagome} out of plane below 175~K. Since the samples are cooled down from room 
temperature in zero magnetic field, the initial state of a macroscopic Co$_3$Sn$_2$S$_2$  flake  is 
magnetically disordered one, e.g., due to magnetic domains. The size of these domains is typically around the 
order of a micrometer~\cite{domains} in Co$_3$Sn$_2$S$_2$, which is much smaller than  the distance between 
the indium leads in our samples. To obtain a definite AHE state of  magnetically ordered WSM, the 
magnetization procedure is performed: an  external magnetic field is swept slowly from -1.5~T to +1.5~T, both
limits are far above the switching positions in Fig.~\ref{cosns_sample}~(b). Afterward, the external field 
goes down to zero. 

Examples of $dV/dI(I)$ characteristics are shown in Fig.~\ref{IV} (a) and (b) before and after the 
magnetization procedure, respectively, for the same sample in a single cooling cycle.

Before magnetization, the curves demonstrate well known Andreev behavior. Since Andreev reflection allows 
subgap transport of Cooper pairs,  it appears experimentally as the resistance drop for voltages within the 
superconducting gap~\cite{tinkham}.  As it can be seen in Fig.~\ref{IV} (a), differential resistance is 
diminished within $\approx\pm 0.5$~mA bias interval in respect to the normal resistance value 
$\approx 0.5\Omega$.  The superconducting gap can be estimated from the width of this region as 
0.5~mA$\times$~0.5~$\Omega \approx 0.25$~meV. Since the bulk indium is known~\cite{indium} to have the 
$0.5$~meV gap, the obtained value is quite reasonable for the indium film on a top of a ferromagnet. 
Temperature has low effect on  $dV/dI(I)$ curves, even at 40~mK the  minimal resistance is not below one half
of the normal value, see Fig.~\ref{IV} (a). 

The magnetization procedure changes the $dV/dI(I)$ curves dramatically, see Fig.~\ref{IV} (b): the zero-bias resistance value  drops to zero. At low temperature 
of 40~mK, we observe a definite zero-resistance state in a wide current region, which qualitatively resembles 
the Josephson effect~\cite{tinkham}.  This behavior has been checked to be independent of  the value and  sign
of the magnetization field.

\begin{figure}
\includegraphics[width=\columnwidth]{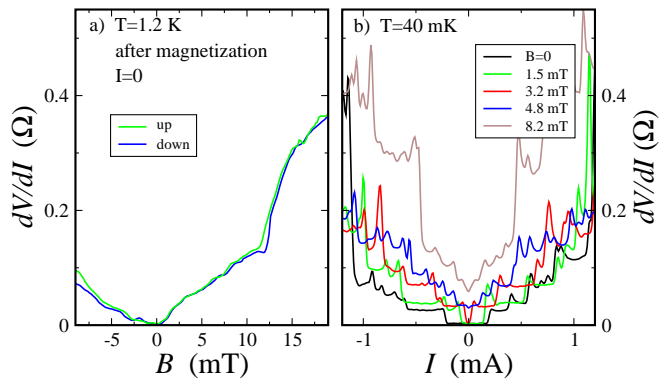}
\caption{(Color online) Suppression of the zero-resistance state by magnetic field. (a) Differential resistance $dV/dI$ at zero dc current as a function of magnetic field. A finite zero-resistance region can  be seen even at highest, 1.2~K, temperature. Two curves correspond to the opposite field sweep directions, they do not strictly coincide  in the normal state. (b)  Examples of $dV/dI(I)$ characteristics for different magnetic fields at 40~mK. The zero-resistance region  survives above 3.2~mT at this temperature. The magnetic field is perpendicular to the flake's plane. The sample is the same as in Fig.~\protect\ref{IV} (after  magnetization).}
\label{IVmag}
\end{figure}
    
As it is expected for the Josephson effect, the zero-resistance state can  be suppressed by magnetic 
field. Even at the highest temperature, the junction resistance is zero in a finite, $\pm 1$~mT field interval, see Fig.~\ref{IVmag} (a). This behavior is demonstrated in detail in Fig.~\ref{IVmag} (b) for $dV/dI(I)$ curves at lowest temperature of 40~mK. The zero-resistance state survives up to 3.2~mT. Above 3.2~mT field, $dV/dI(I)$ curves demonstrate usual Andreev behavior, because the indium leads are still superconducting below the critical indium field~\cite{in-current} of about 40~mT.

\begin{figure}
    \includegraphics[width=\columnwidth]{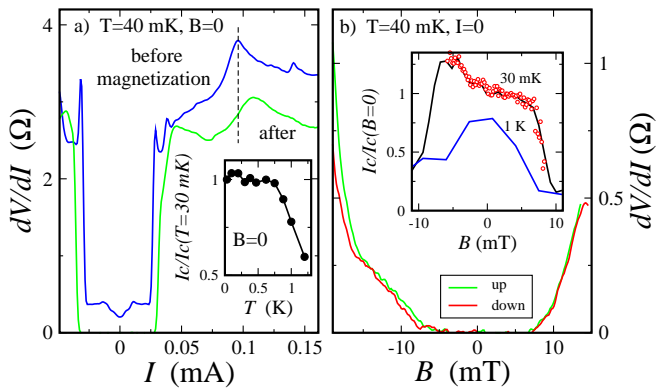}
    \caption{(Color online) Experimental results for a different sample at minimal 40~mK temperature. (a) Formation of the Josephson-type curve (the green one) with well-defined critical current (about 0.025~mA) after sample magnetization. Dashed line indicates the superconductive gap position for the Andreev-like curve (the blue one) without zero-resistance state. Inset demonstrates temperature dependence of the critical current, which is unusual for long diffusive Josephson junctions. (b) Suppression of the zero-resistance state by magnetic field at zero dc current for two  opposite field sweep directions. Inset demonstrates critical current as a function of magnetic field for two different temperatures. At 40~mK, the $I_c(B)$ is  antisymmetric in respect to the zero field, but symmetrically falls to zero at $\pm 7$~mT. The antisymmetry is confirmed by extremely slow field sweep with a low step, as demonstrated by open circles. The  expected $I_c(B)$ symmetry is restored above 1~K. The magnetic field is perpendicular to the flake's plane.}
  \label{jc}
 \end{figure}

Thus, we demonstrate in Figs.~\ref{IV} and \ref{IVmag}, that two superconducting contacts induce Josephson 
current  in an unprecedentedly long  $L=$5~$\mu$m In-Co$_3$Sn$_2$S$_2$-In  junction. The above described 
behavior can be reproduced for different samples, see, e.g., Fig.~\ref{jc}. In this case, the normal $dV/dI$ 
resistance   is one order higher, about 3~Ohm, as depicted in the main field in Fig.~\ref{jc} (a). The 
differential resistance is diminished within $\pm$0.075~mA interval, which gives the same 
0.075~mA$\times$~3~$\Omega \approx 0.25$~meV superconducting gap value. Before sample magnetization,  
$dV/dI$ is always finite even at 40~mK, see Fig.~\ref{jc} (a), while it drops to zero after the magnetization 
procedure.

We  also observe  unusual behavior of the temperature and magnetic field dependences of the critical current 
$I_c$.   Lower currents are more suitable for accurate determination of $I_c$, so the results are presented in the insets to Fig.~\ref{jc} (a) and (b).
 All the experimental points are well reproducible, variation of $I_c$ in different sweeps is below the symbol size in the inset. 

$I_c (T)$ demonstrates weak temperature dependence below 0.75~K, while $I_c$ is diminishing strongly above it
to one half of the initial value at our highest 1.2~K, see the inset to Fig.~\ref{jc} (a). This dependence  
can be crudely extrapolated to $\approx$2~K critical temperature, which well correspond to the $0.25$~meV 
superconducting gap, determined from the Andreev curve in Fig.~\ref{jc} (a).  However, $I_c (T)$  does not demonstrate the conventional for long diffusive SNS junctions exponential decay~\cite{kulik-long,dubos}. The experimental $I_c (T)$ resemble the results for Josephson junctions with spin-flip 
scattering~\cite{dilutedSNS}. 

The zero-resistance state at $I=0$ is suppressed by magnetic field at $\pm$7~mT, see Fig.~\ref{jc} (b). The 
full  $I_c(B)$  pattern  is depicted in the inset to Fig.~\ref{jc} (b). At lowest temperatures,  $I_c(B)$ is 
changing very slowly (within 10\%) until $\pm$7~mT, but falls to zero above this value. The characteristic
feature is that the low-field $I_c(B)$ dependence is antisymmetric in respect to the zero field. This behavior 
is confirmed by extremely slow field sweep with a large amount of points, as demonstrated by open circles in 
the inset. There are also some $I_c(B)$ oscillations within the $\pm$7~mT interval. Both the antisymmetry and 
the oscillations are destroyed by temperature above 0.75~K. The shallow oscillations in $I_c(B)$ could be 
related with usual interference effects~\cite{tinkham}, although the corresponding curve substantially differs
from the standard Fraunhofer pattern~\cite{Barone} and no strong suppression of $I_c(B)$ at $B<\pm$7~mT is observed.

It is important, that both for the resistance at $I=0$ in the main  Fig.~\ref{jc} (b) and for significant 
current $\approx 0.05$~mA  in the inset, the Josephson effect is destroyed under the same magnetic fields 
$\pm$7~mT. Thus, there is no sample overheat in $I_c(B,T)$ measurements, so the unusual $I_c(T)$ temperature  
dependence is correct in the inset to Fig.~\ref{jc} (a).  

\section{Discussion}

As a result, we demonstrate Josephson current through the magnetically ordered $5 \mu$m long 
In-Co$_3$Sn$_2$S$_2$-In junctions, where Co$_3$Sn$_2$S$_2$ flake is in a definite AHE state. On the other 
hand, no supercurrent can be observed in these junctions, when Co$_3$Sn$_2$S$_2$ is in the disordered magnetic
state. This effect is well reproducible for different samples, see Figs.~\ref{IV} and \ref{jc} (a). 

First of all, we should exclude possible fabrication defects, like unintentional shunt connections in the junction's plane. 
The crucial argument against the presence of any leakage is a pronounced dependence of the Josephson 
effect on the WSM magnetization. For conventional superconductors like indium, the proximity to a 
ferromagnet weakens superconductivity. Just opposite to the known indium behavior, magnetically ordered state of 
Co$_3$Sn$_2$S$_2$ enhances the Josephson coupling. Also, the thickness of the indium film is chosen to be much smaller than the leads separation (100~nm$<< 5\mu$m)
to avoid shorting of In leads when pressing the Co$_3$Sn$_2$S$_2$ flake.

Thus, one can be sure that the observed Josephson current flows through the proximity-influenced magnetically 
ordered Co$_3$Sn$_2$S$_2$ WSM. 

The surface states transport can be a key point to explain our experimental results. In the case of 
Co$_3$Sn$_2$S$_2$, Fermi-arc surface states have been directly identified by scanning tunneling
spectroscopy~\cite{kagome_arcs}. The giant anomalous Hall effect is interpreted in terms of the surface 
states~\cite{armitage}, and we observe the Josephson effect only after magnetization procedure, i.e. for a
well pronounced AHE state, see Fig.~\ref{cosns_sample} (b). Therefore, Weyl surface states exist along the 
macroscopic Co$_3$Sn$_2$S$_2$ flake. Their proximity-induced pairing results in the effective 
supercurrent-carrying channels. The important role of the chiral surface channels in the Josephson transport 
studied is also supported by their topological protection and the lateral geometry of the junction.

The magnetic domain structure in Co$_3$Sn$_2$S$_2$ flake can have a substantial influence on the junction's 
properties. When the domain magnetization in the Co$_3$Sn$_2$S$_2$ sample is disordered just after the sample
cooling, the chirality of Weyl nodes can switch across the magnetic domain walls \cite{armitage} and a 
continuous surface state can only appear along a single magnetic domain, which size is  around the order of a
micrometer~\cite{domains}. Such a disordered magnetic structure should produce significant disordered 
spin-flip processes and can prevent the Josephson effect to develope between the 5~$\mu$m spaced 
indium leads, allowing only the observation of Andreev reflection in the proximity of each of the leads, in 
accordance with our measurements for samples before the magnetization procedure.

Since the Weyl surface states are spin-polarized~\cite{armitage,kagome_arcs,cosns_surf} and 
Co$_3$Sn$_2$S$_2$ itself is a half-metal \cite{cosns_surf}, one can expect a triplet supercurrent~\cite{triplet-theor,review-triplet,triplet-exp} through Co$_3$Sn$_2$S$_2$. When singlet Cooper pairs from the superconductor are converted into triplet pairs within the 
spin-polarized material, a long-range proximity effect, as known, can take place: while the singlet component 
penetrates into the ferromagnet over a short length~\cite{singlet-review} $\xi_h=(\hbar D/E_{ex})^{1/2}$ ( $E_{ex}$ is the 
exchange energy and $D$ is the diffusion coefficient), the triplet component penetrates over a much longer length 
$(\hbar D/k_{B}T)^{1/2}$, which is of the same order as that for the penetration of the superconducting pairs into
a normal metal~\cite{triplet-theor}. This conclusion prohibits, in particular, the conventional long-range 
s-wave spin-singlet pairing in Co$_3$Sn$_2$S$_2$ flake, which occurs in In and cannot be present in the 
magnetic Weyl semimetal, as also follows from the chirality blockade arguments \cite{blockade}. Since the 
exchange field and spin-orbit coupling are jointly present in Weyl semimetals, the singlet-triplet conversion 
at its interface does not require additional magnetic 
inhomogeneities~\cite{spin-orbit-triplet1,spin-orbit-triplet2}.  

The triplet supercurrent can be responsible for the low-field $I_c(B)$ antisymmetry in the inset 
to Fig.~\ref{jc} (b). While in topological insulators the surface conducting channels have been experimentally identified as 
dominating in establishing the Josephson coupling~\cite{hyeong}, the Josephson current in Weyl semimetals can
be transferred generally via both the surface and the bulk channels~\cite{dutta}. The low-field variation of $I_c(B)$ reflects the magnetization dynamics in 
the bulk Co$_3$Sn$_2$S$_2$, since in the inset to Fig.~\ref{jc} (b) $I_c$ is smaller for positive fields and 
for positively magnetized flake. As opposed to the low-field region, the fields exceeding  $B_c\approx \pm$7~mT can be considered as destroying the superconducting pairing via the surface channels, where the topologically protected magnetic ordering is not sensitive to the lower values of the external 
field. We wish to mention that uneven indium contact spacing may distort  the behavior of of $I_c(B)$. For example, it (partly)  suppresses the  Fraunhofer pattern~\cite{Barone}. However, it can not make $I_c(B)$ to be sensitive to the sign of the magnetic field, which we observe as the low-field antisymmetry.

\section{Conclusion}

As a conclusion, we experimentally study lateral electron transport between two 5~$\mu$m spaced superconducting indium 
leads on a top of magnetic Weyl semimetal Co$_3$Sn$_2$S$_2$. For the disordered magnetic state of 
Co$_3$Sn$_2$S$_2$ crystal, we only observe the Andreev reflection in the proximity of each of the leads, which
is indicative of highly transparent In-Co$_3$Sn$_2$S$_2$ interfaces. If the sample is homogeneously 
magnetized, it demonstrates well-developed anomalous Hall effect state. In this regime we find the
Josephson current that takes place even for 5~$\mu$m long junctions and shows the unusual magnetic field and 
temperature dependencies. As a possible reason for the results obtained, we discuss the contribution
to the proximity-induced spin-triplet Josephson current from the topologically
protected Fermi-arc states on the surface of Co$_3$Sn$_2$S$_2$.

\acknowledgments
We wish to thank V.T. Dolgopolov for fruitful discussions, and S.V~Simonov for X-ray sample characterization.
We gratefully acknowledge financial support partially by the RFBR  (project No.~19-02-00203), RAS, and RF
State task.

\end{document}